\documentclass[aps,prb,floatfix,showpacs,twocolumn,superscriptaddress]{revtex4}
\usepackage[english]{babel}
\usepackage{graphicx}
\usepackage{dcolumn}
\usepackage{amsmath}

\newcommand{\vs}{{\it vs.}}

\newcommand{\ie}{i.e.}

\newcommand{\alphaETI}{$\alpha$-(BEDT-TTF)$_2$I$_3$}

\begin{document}

\title{Cooperative dynamics in charge-ordered state of $\alpha$-(BEDT-TTF)$_2$I$_3$}

\author{T.\ Ivek}
\email{tivek@ifs.hr}
\homepage{http://real-science.ifs.hr/}
\affiliation{Institut za fiziku, P.O.Box 304, HR-10001 Zagreb, Croatia}
\affiliation{Physikalisches Institut, Universit\"at Stuttgart, Pfaffenwaldring 57, D-70550 Stuttgart, Germany}
\author{I.\ Kova\v{c}evi\'{c}}
\affiliation{Institut za fiziku, P.O.Box 304, HR-10001 Zagreb, Croatia}
\author{M.\ Pinteri\'{c}}
\affiliation{Institut za fiziku, P.O.Box 304, HR-10001 Zagreb, Croatia}
\affiliation{Fakulteta za gradbeni\v{s}tvo, Univerza v Mariboru, Smetanova 17, 2000 Maribor, Slovenia}
\author{B.\ Korin-Hamzi\'{c}}
\affiliation{Institut za fiziku, P.O.Box 304, HR-10001 Zagreb, Croatia}
\author{S.\ Tomi\'{c}}
\affiliation{Institut za fiziku, P.O.Box 304, HR-10001 Zagreb, Croatia}
\author{T.\ Knoblauch}
\affiliation{Physikalisches Institut, Universit\"at Stuttgart, Pfaffenwaldring 57, D-70550 Stuttgart, Germany}
\author{D.\ Schweitzer}
\affiliation{Physikalisches Institut, Universit\"at Stuttgart, Pfaffenwaldring 57, D-70550 Stuttgart, Germany}
\author{M.\ Dressel}
\affiliation{Physikalisches Institut, Universit\"at Stuttgart, Pfaffenwaldring 57, D-70550 Stuttgart, Germany}

\date{\today}

\begin{abstract}
Electric-field-dependent pulse measurements are reported in the charge-ordered state of \alphaETI{}. At low electric fields up to about 50\,V/cm only negligible deviations from Ohmic behavior can be identified with no threshold field. At larger electric fields and up to about 100\,V/cm a reproducible negative differential resistance is observed with a significant change in shape of the measured resistivity in time. These changes critically depend whether constant voltage or constant current is applied to the single crystal. At high enough electric fields the resistance displays a dramatic drop down to metallic values and relaxes subsequently in a single-exponential manner to its low-field steady-state value. We argue that such an electric-field induced negative differential resistance and switching to transient states are fingerprints of cooperative domain-wall dynamics inherent to two-dimensional bond-charge density wave with ferroelectric-like nature.
\end{abstract}


% 71.27.+a
% Strongly correlated electron systems; heavy fermions
%
% 71.45.Lr
% Charge-density-wave systems
%
% 71.30.+h
% Metal-insulator transitions and other electronic transitions
%
% 71.45.-d
% Collective effects
%

\maketitle

\section{Introduction}

Diverse charge structures and complex nonlinear dynamics in anisotropic condensed matter systems with electron-phonon couplings and strong Coulomb interactions have been in focus of intense scientific activity in recent years. Conventional charge and spin density waves (CDW/SDW) previously found in highly anisotropic inorganic and organic systems with prevailing electron-phonon (electron-electron) coupling feature the collective phason mode with well established fingerprints in applied ac and dc electric fields.\cite{Monceau85,Gruener85,Gruener88,Tomic91} Since the phason is pinned to random impurities or the commensurate lattice, it can be observed as a narrow resonance at terahertz frequencies and as a broad loss peak in radio-frequency range. Other famous fingerprints of the phason are the nonlinear conductivity above a finite threshold field and narrow-band noise due to sliding. On the other hand, in the presence of strong onsite and intersite Coulomb repulsion a wealth of broken symmetry insulating ground states with spatially inhomogeneous charge (spin) structures is found in various transition metal oxides and quasi-two-dimensional (quasi-2D) organics.\cite{Fulde93,Seo06} In particular, organics with rich variety of lattice structures provide a nice playground to study electronic phases ranging from localized Wigner-crystal-like charge (spin) to delocalized CDW-like (SDW) modulation of charge (spin) densities. However, there are no well-developed theoretical models about collective excitations in these phases and how they should respond to applied ac and dc electric fields.

Under applied ac electric fields, anisotropic phason-like dispersions were found similar to those of conventional CDWs, while at high dc electric fields contradictory results were reported ranging from negligibly small nonlinearities with no threshold field to the well defined effects due to sliding similar to conventional CDWs.\cite{Vuletic2006, Ivek2008, Blumberg2002} At much higher fields the giant nonlinear conductivity and negative differential resistance (NDR) were reported\cite{Yamanouchi99,Sawano2005,Niizeki2008,Ozawa2009,Mori2009} which were very different from those known in conventional CDWs. Distinct mechanisms to explain these remarkable nonlinear effects have been suggested ranging from melting of charge order in $\theta$-(BEDT-TTF)$_2$Cs$M'$(SCN)$_4$ ($M'=\textrm{Zn, Co}$) compounds \cite{Sawano2005} to dielectric breakdown and depinning and sliding of holes within charge stripes in La$_{2-x}$Sr$_x$NiO$_{4+\delta}$.\cite{Yamanouchi99} In the TMET-TTP organic system the electrothermal model originally developed to account for the hot electron dynamics was used to reproduce the measured nonlinear conductivity which featured NDR with no switching.\cite{Mori2009} On the other hand, in $\theta$-(BEDT-TTF)$_2$$M$Zn(SCN)$_4$ ($M=\textrm{Cs, Rb}$) and \alphaETI{} the $I-V$ characteristics were found to follow a power-law behavior $I \propto V^\alpha$ in the range of extremely low currents (lower than 10\,nA) and low temperatures (lower than 60\,K) and were explained within the excitonic model with a long-range Coulomb interaction.\cite{Takahide10,Kodama12} Further, an analogy to SDW sliding has been invoked to explain unusual nonlinear effects in the case of antiferromagnetic phase of localized spins in MDT-TS organic compound.\cite{Mori2008} Finally, when the constant voltage is applied to the sample, current oscillations with very low fundamental frequencies (10--100\,Hz) have been observed in the NDR region of several quasi-2D organics and interpreted as the switching frequency between low and high current states in analogy to the well-known Gunn effect observed in two-valley semiconductors.\cite{Conwell67} The cause of all these remarkable nonlinear effects, although appearing as a common phenomenon in many strongly correlated insulating systems, is still unclear and the proposed interpretations are highly controversial. 

\alphaETI{} represents one of the most prominent and best studied quasi-2D organic metals with charge order. Synchrotron x-ray diffraction, NMR, Raman and infrared vibrational measurements (Ref.\ \onlinecite{Ivek2011} and references therein) demonstrate that \alphaETI{} should be considered as the model case of the 2D bond-charge density wave with concomitant formation of a pattern of structural twin domains.\cite{Clay02} These results at the same time indicate that the description of charge order as a horizontal stripe of localized charges is over-simplified and that the modulation of bonds in two almost diagonal in-plane directions needs also to be taken into account. Recently, on the basis of the anisotropic electrodynamic response it was shown that collective excitations of such a charge order possess a complex anisotropic dispersion along diagonal as well as long two crystallographic in-plane directions: a broad screened relaxation due to long wavelength phason-like excitations and a temperature-independent relaxation with small amplitude due to the motion of domain-wall pairs between twinned domains in the charge order texture.\cite{Ivek2010} If this interpretation is valid, the dynamical properties of these collective excitations should also manifest themselves as a characteristic nonlinear response in high dc electric fields. Early experiments by one of us first observed nonlinear effects in dc conductivity at low fields.\cite{Dressel94} Further nonlinear conductivity measurements along three crystallographic directions have shown NDR and voltage oscillations.\cite{Tamura2010} The authors point out that the current-voltage characteristics measured in two-probe configuration by constant current and constant voltage method were essentially the same and oscillations could be detected only for sufficiently short single crystals. The picture of pure horizontal stripe charge order was invoked to interpret these results as the evidence for charge order depinning and sliding.

In this communication we attempt to clarify these important issues concerning the nature and origin of nonlinearities in the charge-ordered ground state of \alphaETI{}. We approach the charge order as a 2D bond-charge density wave developed in the pattern of ferroelectric-like twin domains. Our results give evidence that long-wavelength collective excitations of this charge order, while possessing phason-like dispersion, do not show the nonlinear response in high dc electric fields established for conventional CDWs. Rather, short-wavelength excitations with domain-wall-like dispersion give rise to a negative differential resistance at high dc electric fields and reversible switching to transient high-conducting states. 

\begin{figure} %FIG1
\includegraphics[clip,width=1.0\columnwidth]{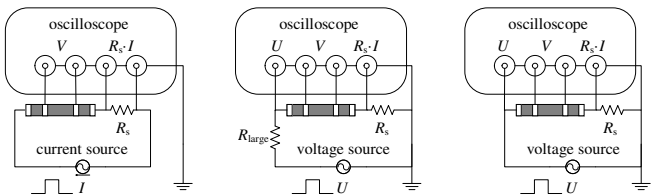}
\caption{\label{fig:schematics} Schematic diagrams of the pulse measurement circuits employed for the constant-current mode (left and middle panels) and for the constant-voltage mode (right panel).}
\end{figure}

\section{Experimental}

The measurements were performed on six single crystals of varying lengths (0.2--2\,mm) along three in-plane directions: in the $a$ and $b$ crystallographic axes and in diagonal [110] direction. The sample selected for a detailed study for $\mathbf{E}\parallel a$ had a cross-sectional area of 0.0032\,mm$^2$, and the distance between current and voltage contacts was 0.37 and 0.11\,mm, respectively. The electric transport in the Ohmic regime was measured between room temperature and 20\,K as described in Ref.\ \onlinecite{Ivek2011}; qualitatively the same behavior was found with no hysteresis in the phase transition temperature region.

At high fields the current-voltage characteristics were obtained by pulse measurements in the charge-ordered state between 100 and 135\,K, a temperature range where systematic errors were excluded due to sample resistances much smaller than input impedance of instrumentation. Mostly a four-point configuration was used in order to eliminate any possible contact influence. The experiments were carried out using two measurement modes as shown in Fig.\ \ref{fig:schematics}. In the constant-current mode, a circuit with Keithley 6221 current-source was used (Fig.\ \ref{fig:schematics}, left panel). In order to minimize Joule heating we applied rectangular pulses, typically 1-5\,ms wide with a settling time of about 5\,$\mu$s and the repetition time of 100--500\,ms. The sample resistance was measured by monitoring the voltage drop $V$ at the sample and the current $I$ through load resistor $R_\mathrm{s}$ by a Tektronix TDS3014C digital storage oscilloscope. We also measured the total source voltage $U$. The response was typically sampled 64 times and averaged. Interestingly, out of six samples the two with a distance between current contacts smaller than 0.4\,mm showed real-time voltage oscillations, which is in accord with previous findings.\cite{Wakita2010,Tamura2010} However, the current pulses turned out to feature oscillations as well, indicating that the current source did not succeed in regulating a constant current within the pulse.\cite{note1} The spurious origin of oscillations was confirmed by the experiments using another circuit for the constant-current mode in which the current was fixed by means of a large resistor in series with an HP214B pulse generator (Fig.\ \ref{fig:schematics}, middle panel). These experiments as well as experiments done in the constant voltage mode never did exhibit any oscillations, in any of its various configurations.

In the circuit for the constant-voltage mode a Tabor 8023 signal generator (Fig.\ \ref{fig:schematics}, right panel) provided voltage pulses (typical width of 1--5\,ms, repetition time 100--500\,ms), and the voltage drop and current through the sample is again measured by an oscilloscope.

\begin{figure*} %FIG2
\includegraphics[clip,width=1.0\linewidth]{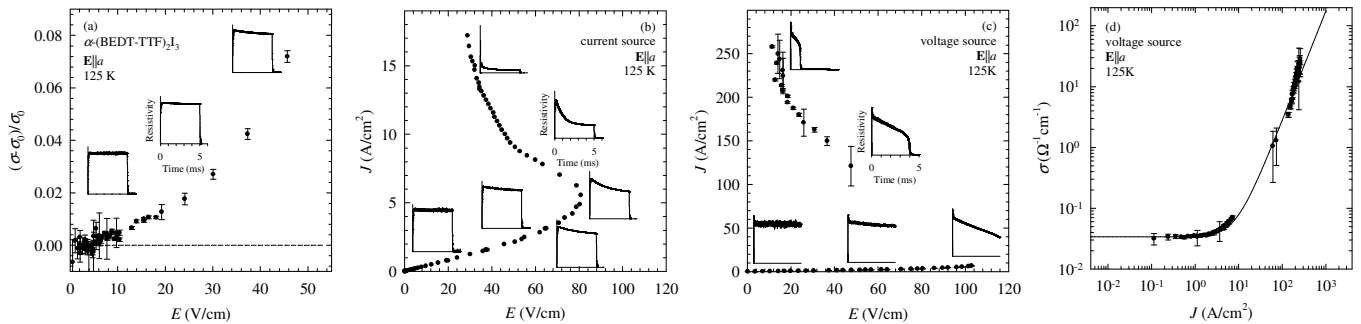}
\caption{\label{fig:125K}(a) Low- and (b, c, d) high-electric field behavior of \alphaETI{}, $\mathbf{E}\parallel a$, at a representative temperature 125\,K. (a) Conductivity increase normalized to its Ohmic value \vs{}\ electric field.
(b) Current density \vs{}\ electric field with corresponding resistivity \vs{}\ time plots measured in the constant-current and (c) constant-voltage mode. The current density and electric field are evaluated as an average of 4--5\,ms.
(d) Conductivity \vs{}\ current density measured in the constant-voltage mode. The full line is a fit to the power law $\sigma(J,T) = \sigma_0(T) + A J^n$ with $\sigma_0= 0.034$\,$\Omega^{-1}$cm$^{-1}$ and $n = 1.8\pm 0.2$. Constant-current mode data follow the same expression with $\sigma_0= 0.04$\,$\Omega^{-1}$cm$^{-1}$ and $n = 2.1\pm 0.2$.}
\end{figure*}

We have additionally recorded time-resolved infrared reflectivity spectra, $\mathbf{E}\parallel a$ and $b$, during the above-described voltage pulses. These preliminary measurements with a time resolution of 25\,$\mu$s have been performed in the frequency range 800--4000\,cm$^{-1}$ using a Bruker Vertex 80v spectrometer in step-scan mode as a trigger for the voltage-pulse generator and a Hyperion IR Microscope with a measurement spot between voltage contacts.\cite{Knoblauch12}

\section{Results and Discussion}

All measured temperatures show qualitatively the same transport behavior. A typical non-linear conductivity for fields up to 50\,V/cm for $\mathbf{E}\parallel a$ measured in constant-current mode is shown in Fig.\ \ref{fig:125K}(a). The obtained behavior is in accord with that of the constant-voltage mode measurement. It is evident that the threshold field for sliding cannot be clearly defined.\cite{note2} Qualitatively similar results were obtained for the conductivity data along the $b$-axis and diagonal [110] direction as well. All orientations show an almost negligible nonlinearity (a few percent) if compared with the standard CDW nonlinear effects which amount to at least several tens of percent at twice the threshold field.\cite{Monceau85,Gruener85,Gruener88} Pulse shapes show only a small deviation from rectangular shape.

The sample response drastically changes in higher fields [Figs.\ \ref{fig:125K}(b) and (c)]: the current density $J$ against the electric field $E$ curve deviates from the Ohmic-like behavior at fields higher than 50\,V/cm and eventually shows a negative resistance behavior (NDR).\cite{Tamura2010} Our results clearly demonstrate that the observed NDR which sets in above the threshold field $E_\mathrm{th}$ is accompanied with a significant change in shape of the measured resistivity $R$ in time $t$ which critically depends on the applied measurement mode.\cite{note3} When the constant-current mode is used, $R$--$t$ curves show only smooth changes within pulse [Fig.\ \ref{fig:125K}(b)]; conversely when the constant-voltage mode is used, $R$--$t$ curves show a discontinuous jump to a low-resistance state [Fig.\ \ref{fig:125K}(c)] and an overall drop to low fields in $E$--$J$ plots happens at significantly higher current densities. Delay times become longer for fields close to threshold indicating the field-dependent formation of the connected metallic-like regions. The NDR behavior becomes more prominent and the threshold field $E_\mathrm{th}$ increases as the temperature lowers (not shown). Unfortunately, due to a rather restricted temperature range we cannot distinguish between conventional CDWs, for which at low temperatures theory predicts $E_\mathrm{th}(T) \propto \exp(-T/T_0^\prime)$,\cite{Maki86} and dielectric breakdown associated with the charge carrier avalanche process which experimentally follows the $E_\mathrm{th}(T) \propto \exp(T_0^{\prime\prime} /T)$ law.\cite{Taguchi2000} In the former case $T_0^\prime$ corresponds to the strength of the pinning potential, and we get a value of 5\,K. This is close to that found for charge stripes (2.5--15\,K), and somewhat smaller than the values typically found in conventional CDWs (20--100\,K). On the other hand, our data nicely follow the phenomenological formula for current-dependent conductivity expected in the presence of avalanche process of current, \ie{}, charge carriers, $\sigma(J,T) = \sigma_0 (T) + A J^n$, where the first term is Ohmic conductivity and $n$ in the second term does not depend on temperature [Fig.\ \ref{fig:125K} (d)]. 

\begin{figure} %FIG3
\includegraphics[clip,width=1.0\columnwidth]{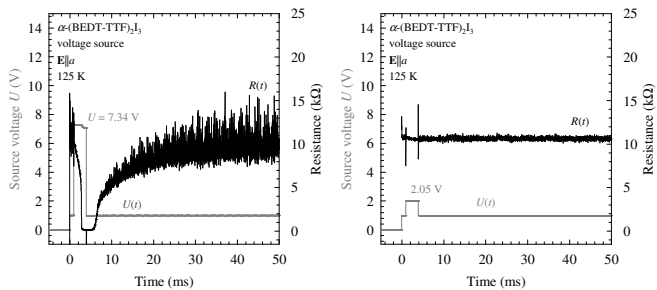}
\caption{\label{fig:memory} Effects of two extreme conditioning pulses with total source voltage $U$ on \alphaETI{}, $\mathbf{E}\parallel a$, at a representative temperature 125\,K measured in constant-voltage mode. At sufficiently high electric fields within the conditioning pulse the sample becomes metallic and in time decays back to the high-resistance state (left panel). A low-field conditioning pulse keeps the resistance unchanged (right panel).}
\end{figure}

We also characterize the switching from a high to low-resistance state as detected in the constant-voltage measurement mode, and to test for a possible memory effects we have applied the following procedure (Fig.\ \ref{fig:memory}). First, a very short (1\,ms) low-field pulse was applied to measure the sample resistance; it was followed by an adjustable high-field conditioning pulse (3\,ms), and in the end a long, low-field probing pulse was applied to track time-dependent recovery of sample resistance (up to 100\,ms). Our results demonstrate that the low-resistance states achieved by switching are transient in nature: once the conditioning pulse with a voltage high enough to induce switching into a low-resistance state is turned off, the memory keeps the sample metallic-like during 2\,ms and subsequently the sample's resistance decays back in a single-exponential manner to its steady-state value. The switching to a low-resistance state and back is a highly reproducible phenomenon. Additionally, after the voltage is applied, the switching occurs at a characteristic delay time which becomes shorter either for larger voltage amplitudes at a given temperature (0.8--17\,ms at 125\,K) or at higher temperatures for a given voltage amplitude (not shown).

\begin{figure} %FIG4
\includegraphics[clip,width=0.44\columnwidth]{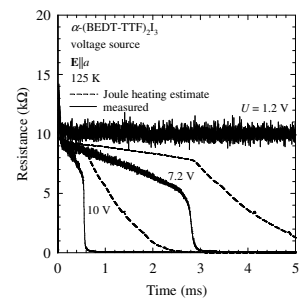}
\caption{\label{fig:analysis} Comparison between measured resistance \vs{}\ time (full lines) and an estimate (dashed lines) based on Joule heating and the dc curve, for three representative pulses with total source voltage $U$ applied to \alphaETI{} along the $a$-axis at 125\,K.}
\end{figure}

Although such a huge (by more than a factor of 100, see Fig.\ \ref{fig:memory}, left panel) and steep reduction of resistance under high electric field can hardly be attributed to Joule heating, the decay time dependency motivated us to verify the extent of its influence. The resistance at 125\,K before the switching was 10\,k$\Omega$. The switching to low-resistance state of about 40\,$\Omega$ was observed with delay time of 3\,ms upon the application of 7.34\,V. The total Joule heat generated in the sample prior to switching was calculated to be less than 1\,$\mu$J. Using the measured value of heat capacity (575\,J/molK at 125\,K)\cite{Fortune91} and the sample volume, the upper bound of temperature rise due to Joule heating would be less than 1\,K with the assumption of homogenous temperature. Fig.\ \ref{fig:analysis} shows a comparison between measured resistance \vs{}\ time as opposed to estimates based on Joule heating and dc resistivity curve. A similar data analysis was done for the constant-current mode experiments where only a smooth decrease of the resistance with no delay was detected (not shown). The results verified that even in this measurement mode the resistance fell faster than the Joule heating estimate, which implies that heating cannot be the sole cause of the observed effect. Preliminary time-resolved infrared spectroscopy experiments also indicate that the sample does not undergo a transition into the high-temperature metallic phase as it switches to a low-resistance state during a voltage pulse. To elaborate, an \alphaETI{} heated through the metal-to-insulator transition would show characteristic changes in the $\mathbf{E}\parallel a$, $b$ infrared spectra before and after transition.\cite{Ivek2011} In our time-resolved infrared experiment these changes were notably absent, indeed there are only negligible differences to the insulating-phase spectra through the whole duration of a high-field switching pulse.\cite{Knoblauch12} Therefore, we are confident that the NDR with switching and the low-resistance state should be an intrinsic, although transient, state of electron system in the charge-ordered state of \alphaETI{}.

Nonlinear conductivity effects similar to the ones in \alphaETI{} have also been observed in diverse quasi-1D strongly correlated inorganic and organic insulating systems. Arguments were given that the observed dynamics should be collective in nature, analogous to the case of conventional CDWs. First, the decrease of $E_\mathrm{th}$ with increasing temperature of the form $E_\mathrm{th}(T)/E_\mathrm{th}(0) = \exp(-T/T_0^\prime)$ has been also found in CDWs and ascribed to the reduction of pinning by thermal fluctuations of the phase of the CDW order parameter.\cite{Maki86} Second, the magnitude of threshold fields, although several orders of magnitude larger than needed to depin a CDW ($\sim 0.01$--1\,V/cm), is at the same time much smaller (again by several orders of magnitude) than typical breakdown fields in band insulators ($10^7$\,V/cm). Third, delayed switching is an effect also found in conventional CDWs\cite{Mihaly84} and it was consistently explained as a result of coupling among CDW domains to form a coherent current-carrying state. As a rule the nonlinear contribution to conductivity in conventional CDWs does not follow the $J^n$ power law. To the best of our knowledge, NbSe$_3$ is the only conventional CDW system which shows an NDR,\cite{Hall84} however the effect is very small and dwarfed by the drop in resistivity of \alphaETI{}.

In \alphaETI{} the values of threshold fields are more than one order of magnitude smaller than the ones found in transition-metal Mott insulators and mixed-stack organic systems.\cite{Taguchi2000,Tokura88,Iwasa89} Further, the delay times (0.01--10\,ms) are 10 times longer than known from standard CDW system (0.001--0.1\,ms),\cite{Mihaly84} but more than 100 times shorter than the ones detected in transition metal compounds with stripe phases.\cite{Yamanouchi99} Delayed switching in conventional CDWs was studied in relation to sign memory charging and was monitored as transient voltage responses to current pulses with different histories.\cite{Mihaly84} In these experiments, as in many others, periodic voltage oscillations were clearly resolved. Memory effects were attributed to the redistribution of CDW deformations and the appearance of voltage oscillations to the interaction of the sliding CDW with impurities. Voltage-mode experiments on K$_{0.3}$MoO$_3$ at very low temperatures where electron screening is absent have shown behavior atypical for conventional CDWs: strong nonlinearities, delayed current switching and oscillating current response were detected above an extremely high electric field of about 100\,V/cm.\cite{Maeda85} As far as the voltage oscillations are concerned our results clearly show that the voltage oscillations in our samples, when they appeared, are only an artifact of the measurement circuit and are in sharp contrast to previously published two-contact measurements.\cite{Tamura2010} We are led to conclude that a more systematic investigation needs to be done on single crystals of different geometries, from different syntheses and using different constant-current circuits in order to reconcile the existing observations.

Therefore, we consider that delayed switching detected in the NDR regime as a transient current response and the power-law current-dependent conductivity seem to be a somewhat different phenomenon than the one in most conventional CDWs. However, the very reason why the charge response at high electric fields is so different may be due to the complex structure of the CO in \alphaETI{}. Namely, the unit cell consists of four crystallographically inequivalent molecules placed in different environments, which intuitively should prevent sliding in high fields as found in conventional CDWs. The 1100 modulation of site charges accompanied by bond tetramerization along the two diagonal in-plane directions appears not to be the only significant pattern of the CO.  Indeed, the symmetry breaking at the CO phase transition is associated with a 1010 modulation and bond dimerization within the stack of A molecules perpendicular to stripes. The latter change implies the loss of inversion centers between A molecules which consequently become non-equivalent and thus are responsible for the ferroelectric-like nature of the CO.\cite{Kakiuchi07} Recently it was suggested on the basis of DFT calculations that the instability which leads to this CO and associated ferroelectricity is not purely Coulumb-driven within the molecular subsystem, but that it is also amplified by anion interactions.\cite{Alemany12}

On the other hand, the observed behavior is rather similar to the one of strongly correlated La\-Sr\-Ni\-O with charge stripes,\cite{Yamanouchi99} the SrCuO Mott insulators,\cite{Taguchi2000} as well as in organic systems.\cite{Niizeki2008} In the La\-Sr\-Ni\-O extremely long delay times near threshold field were taken as an indication of the field-dependent formation of conductive regions and the onset of NDR was suggested to be a sign of a dielectric breakdown and concomitant collective motion of holes within charge stripes. The dielectric breakdown was also suggested in an organic conductor based on BEDT-TTF due to a resistance drop by a factor of $10^3$ [Ref.\ \onlinecite{Niizeki2008}], while in a mixed-stack charge-transfer system\cite{Tokura88} the NDR was attributed to either the cooperative motion of charge defects or domain walls inherent to the insulating states of these systems. In all these cases as well as in \alphaETI{} the current-power-law dependent conductivity was found with the exponent $n = 1.8$ implying a self-multiplication avalanche-like process of charge carriers which is induced by current density. Such a process can be also seen as a cooperative percolation process among randomly distributed pre-formed conductive domains switched on at fields above $E_\mathrm{th}$.

In case of switching from insulating to a metallic-like state in \alphaETI{} we have to consider the following possible causes. The first one could be the destruction of charge order by high electric fields as described above for strongly-correlated systems, or as in PrCaMnO.\cite{Asamitsu97} Another option are electron heating effects. Simple estimates indicate that both of these causes can be ruled out. Namely, taking the upper bound for the inelastic electron mean free path of about 1.35\,nm in the temperature range 100--135\,K,\cite{note4} the maximum energy provided by a field of 100\,V/cm is of the order of $13.5\,\mu\textrm{eV}\approx 0.16$\,K. This value is still four times smaller than the charge order energy gap and far below the thermal energy at measured temperatures. Dielectric breakdown can be also ruled out since no oscillatory behavior of current, predicted by theories, was observed in our samples.\cite{Conwell67, Oka2003} It is noteworthy that the avalanche-like increase of conductivity can happen at fields which are still lower than needed for the onset of breakdown. Here we also point out a recent pump-probe experiment which demonstrated a photoinduced transition in \alphaETI{} from insulating to the metallic state.\cite{Iwai} The initial fs-scale pulse destroys the charge order; macroscopic metallic domains form and eventually relax back to the charge-ordered state in ps--ns times. These fast relaxations are in contrast to the millisecond-scale high-conducting state found in our experiments. 

Keeping in mind the above considerations, we suggest the following scenario for the charge response in high dc electric fields of the CO developed in \alphaETI{}.\cite{Ivek2011} The long-wavelength collective excitations with phason-like dispersion are not able to respond to the applied dc electric field due to structural inequivalences of BEDT-TTF molecules within a unit cell.\cite{Alemany12} Rather we propose that the domain-wall pairs formed due to the twinned nature of the charge-ordered phase are responsible for the observed nonlinearities. The observed millisecond scale of response is indeed too slow for phasons; more likely it corresponds to soliton-like excitations. In the applied dc electric field the coupling to the dipole moments of each unit cell breaks the symmetry between two opposed dipole orientations. At high fields the rate of formation of domain wall pairs strongly increases and overcomes their formation due to thermal excitations. Subsequently, the motion of domain wall pairs become increasingly correlated and thus creates growing conduction regions until percolation promotes an NDR. The microscopic nature of these regions remains open, that is whether percolation islands or filament-like regions are formed. It is noteworthy that a qualitatively similar behavior is observed in the constant current mode, albeit without switching and thus significantly weaker. This result implies that the NDR itself is independent of measurement-mode but achieving a prominent effect absolutely requires measurement conditions which allow unlimited current self-multiplication: an uncapped, \ie{}, voltage source. 
 
\section{Conclusion}
In summary, we demonstrate the huge negative differential resistance and switching to transient high-conduction states at high dc electric fields within molecular planes in the charge-ordered state of \alphaETI{}. Conversely, at low dc electric fields only negligibly small nonlinearities without a clear-cut threshold field are observed. We argue that this behavior of dc conductivity is closely related to the origin of complex dielectric response at radio-frequencies. In high-field regime, we suggest that the effects are accounted for by the dynamics of short-wavelength excitations of the two-dimensional bond-charge density wave, \ie{}, domain walls at interfaces of two types of ferroelectric-like domains. On the other hand, in low-field regime, negligibly small effects indicate a largely frozen dynamics of collective long-wavelength (phason-like) excitations whose dispersion at low ac fields we have detected previously as a broad screened relaxation mode. Further experiments are needed to clarify whether such a negative differential resistivity phenomenon is a unique feature of the charge order in \alphaETI{}, or is common to strongly correlated systems close to commensurability.

\begin{acknowledgments}
We thank G.\ Untereiner for sample preparation. ST acknowledges helpful discussions with S.\ Mazumdar and R.\ T.\ Clay. This work was supported by the Croatian Ministry of Science, Education and Sports under Grants No.\ 035-0000000-2836 and by the Deutsche Forschungsgemeinschaft (DFG). TK was supported by the Carl Zeiss Stiftung.
\end{acknowledgments}


\begin{thebibliography}{99}
\bibitem{note1}
This is not surprising since in the presence of an NDR and high enough capacitance in parallel it might be possible to drive the feedback loop of the current source into an oscillating regime.
\bibitem{note2}
A threshold field of about $5\pm 1$\,V/cm may be seen in the lin-log plot; however the log scale can be misleading and is usually not employed.
\bibitem{note3}
The two circuits for constant current mode gave similar $J$--$E$ curves, which excluded the possibility that the current source feedback loop influences the pulse shape in a systematic way.
\bibitem{note4}
The electron mean free path is estimated to be about 1\,nm at room temperature and 1.35\,nm just above the charge-ordered phase transition temperature.
\end{thebibliography}
\end{document}